\documentclass[twocolumn, prl, 10pt, superscriptaddress, footinbib]{revtex4-1}
\usepackage{graphicx, subfigure, amsfonts,amssymb,amsmath, array, gensymb,fontenc,color, lipsum}

\begin{document}

\title{{Electronic transport across quantum dots in graphene nanoribbons: \\ Toward built-in gap-tunable  metal-semiconductor-metal heterojunctions}}
\author{Kristi\={a}ns \v{C}er\c{n}evi\v{c}s}
\affiliation{Institute of Physics, Ecole Polytechnique F\'ed\'erale de Lausanne (EPFL), 1015 Lausanne, Switzerland}
\affiliation{National Centre for Computational Design and Discovery of Novel Materials (MARVEL),  Ecole Polytechnique F\'ed\'erale de Lausanne (EPFL), 1015 Lausanne, Switzerland}
\author{Oleg V.\ Yazyev}
\affiliation{Institute of Physics, Ecole Polytechnique F\'ed\'erale de Lausanne (EPFL), 1015 Lausanne, Switzerland}
\affiliation{National Centre for Computational Design and Discovery of Novel Materials (MARVEL),  Ecole Polytechnique F\'ed\'erale de Lausanne (EPFL), 1015 Lausanne, Switzerland}
\author{Michele Pizzochero}
\email{mpizzochero@g.harvard.edu}
\affiliation{Institute of Physics, Ecole Polytechnique F\'ed\'erale de Lausanne (EPFL), 1015 Lausanne, Switzerland}
\affiliation{National Centre for Computational Design and Discovery of Novel Materials (MARVEL),  Ecole Polytechnique F\'ed\'erale de Lausanne (EPFL), 1015 Lausanne, Switzerland}
\affiliation{School of Engineering and Applied Sciences, Harvard University, Cambridge, MA 02138, USA}

\date{\today}
\begin{abstract}

Success of all-graphene electronics is severely hindered by the challenging realization and subsequent integration of semiconducting channels and metallic contacts. Here, we comprehensively investigate the electronic transport across width-modulated heterojunctions consisting of a graphene quantum dot of varying lengths and widths embedded in a pair of  armchair-edged metallic nanoribbons, of the kind recently fabricated \emph{via} on-surface synthesis. We show that the presence of the quantum dot enables to open a width-dependent transport gap, thereby yielding built-in one-dimensional metal-semiconductor-metal junctions. Furthermore, we find that, in the vicinity of the band edges, the conductance is subject to a smooth transition from an antiresonant to a resonant transport regime upon increasing the channel length. These results are rationalized in terms of a competition between quantum-confinement effects and quantum dot-to-lead coupling. Overall, our work establishes graphene quantum dot nanoarchitectures as appealing platforms to seamlessly integrate gap-tunable semiconducting channels and metallic contacts into an individual nanoribbon, hence realizing self-contained carbon-based electronic devices.

\end{abstract}
\maketitle

\paragraph{Introduction.}
In addition to the unique features arising from its peculiar $\pi$-electron network, which include high structural flexibility, superior mechanical stability, and ambipolar ballistic transport of massless Dirac fermions over microscopic length scales \cite{Novo05a, Meu16, Neto09}, the properties of graphene can be tremendously shaped upon nanostructuring in one-dimensional nanoribbons. This is typically accomplished by means of bottom-up synthetic routes, in which surface-assisted Ullman-type coupling of molecular precursors is exploited to obtain polyphenylene intermediates that are eventually converted into graphene nanoribbons (GNRs) through cyclodehydrogenation \cite{Cai10a}. Importantly, the structure of the atomically precise reaction products is fully defined by the selection of the precursor monomers \cite{Cai10a}. This approach has allowed to manufacture a broad assortment of GNRs featuring diverse widths, edge geometries and chemical compositions \cite{Yano20}, hosting novel physical phenomena that range from magnetism \cite{Sun2020m} to band topology \cite{Cao2017, Groning2018, Rizzo2018}. 

Within this wide spectrum of nanostructures and accompanying properties, armchair graphene nanoribbons (AGNRs) are arguably the most extensively studied. In these systems, quantum confinement effects can open a tunable band-gap in their energy spectrum, hence conferring switching capabilities to otherwise semi-metallic graphene \cite{Llinas2017, ElAbbassi2020, Pizzochero2020}. AGNRs can be grouped into three distinct families, depending on the number of carbon atoms in the direction transverse to the ribbon axis, namely the $3p$, $3p + 1$, and $3p + 2$ classes, with $p$ being a positive integer \cite{Son06a, Yaz13}. While the $3p$ and $3p + 1$ classes exhibit sizable band-gaps that scale inversely with the nanoribbon width \cite{Chen13}, the $3p + 2$ one retains a (quasi-)metallic character \cite{Kimouche2015}. The large degree of control that has been experimentally achieved over the atomic-level features of AGNRs has sparked the opportunity to realize complex one-dimensional heterostructures, including two- \cite{Blan12a, Ma19a}, multiple- \cite{Cai10a}, as well as hetero-terminal junctions \cite{Bron18a, Nguy17, Cai14a}.
 
Of particular interest among this rich variety of carbon-based architectures  is the assembly of width-modulated armchair graphene nanoribbons \cite{Wang2017, Chen15a, Jacobse2017}.   These heterojunctions have been recently fabricated either through lateral fusion of two distinct AGNR segments of different length \emph{via} cross-dehydrogenative coupling \cite{Wang2017, Chen15a} or following a bottom-up approach \cite{Jacobse2017} on Au(111) surface, yielding armchair nanoribbons that are smoothly edge-functionalized with guest graphene quantum-dots of varying length and width. Such nanoarchitectures may hold promise towards graphene electronics by virtue of the tunable band-gap of the constituent AGNR building-blocks \cite{Jacobse2017}. Although graphene-based quantum dots and constrictions have been previously studied \cite{Gonz11, Xiong11, Darancet2009}, a comprehensive picture of the electronic structure of the aforementioned width-modulated armchair graphene nanoribbons is still missing.

In this Rapid Communication, we fill this gap in knowledge by systematically investigating the charge transport across graphene quantum-dots of different geometries embedded in armchair-edged graphene nanoribbons. With the help of atomistic simulations, we achieve a detailed understanding of the interplay between their atomic structure and electronic transport.  Importantly, we establish guidelines to assemble all-graphene electronic devices composed of a semiconducting quantum-dot channel seamlessly contacted to metallic leads. Overall, our findings ignite the potential of as-synthesized graphene quantum-dots to develop complete built-in carbon-based electronic devices in one dimension.

\medskip
\paragraph{Methodology.}  
Our calculations are conducted at the tight-binding level, in which the Hamiltonian $H$ describing the $p_z$ electrons takes the form
\begin{equation}
H = \sum_i \epsilon_i c_i ^\dagger c_i + t \sum_{<i,j>} (c_i ^\dagger c_j +\textnormal{H.c.}),
\label{Ham}
\end{equation} 
with $c_i^\dagger$  ($c_i$) being the  creation (annihilation) operator at the $i$-th lattice site. We neglect the on-site potential $\epsilon_i$ and set the nearest-neighbors hopping integral $t$ to 2.75 eV. The choice of resorting to the tight-binding framework instead  of \emph{e.g.}\ first-principles calculations is dictated by the large number of atoms considered in our realistic models (up to several thousands) and supported by the well-known satisfying description that the Hamiltonian of Eqn.\ (\ref{Ham}) provides as far as the experimentally relevant low-energy physics of graphene nanostructures is concerned \cite{Yaz13}. Admittedly, the adopted theoretical framework neglects possible charging effects.

 \begin{figure}[t]
  \centering
 \includegraphics[width=1\columnwidth]{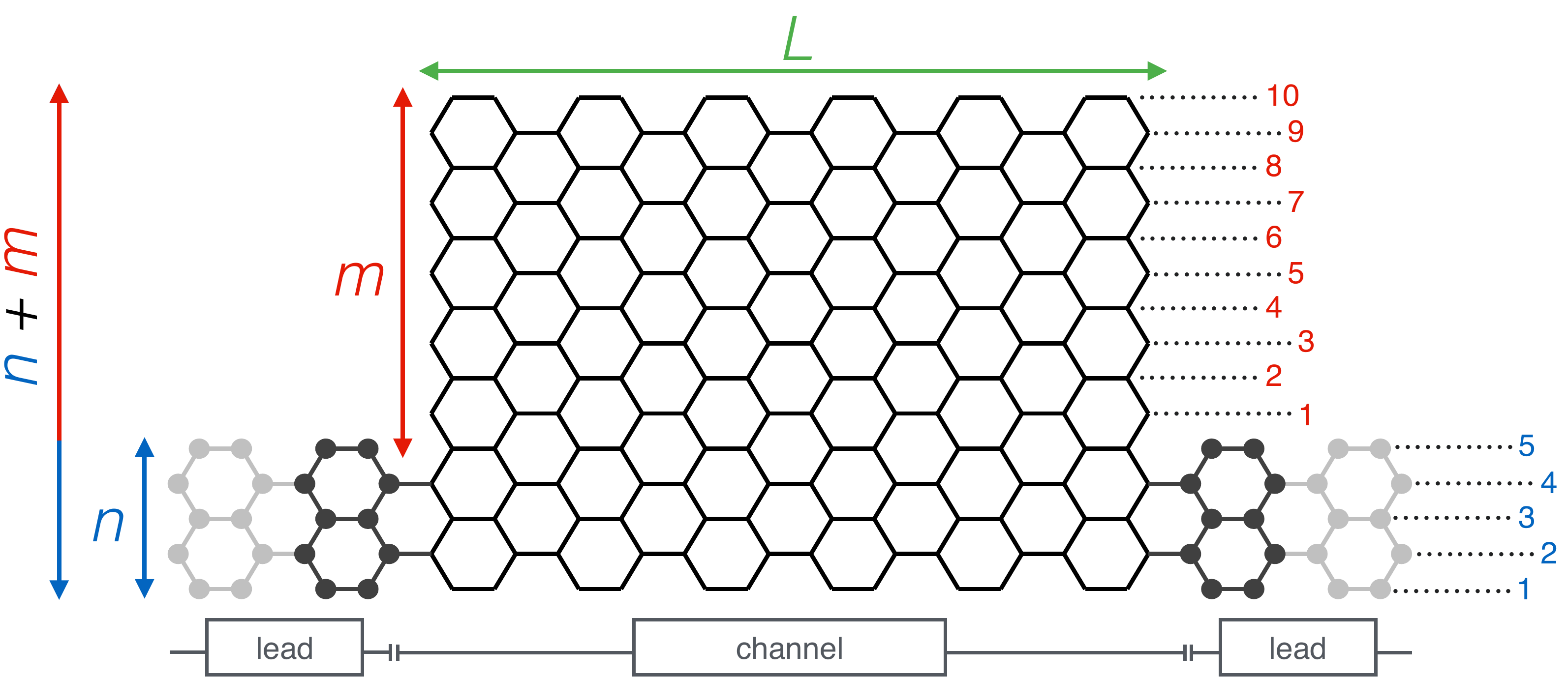}
  \caption{Atomic model of the nanostructure considered in our electron transport calculations, in which a quantum dot of varying width $m$ (with $5 \leq m \leq 13$) and length $L$ (with $0.9$ nm $\leq L \leq 21.3$ nm) acts as a channel region embedded in a pair of metallic $n$-AGNRs leads (with $n$ = 5, 8, 11), giving rise to a $n$-$(m+n)$-$n$-AGNRs  heterojunction. \label{Fig1}}
\end{figure}

Electronic transport properties are next calculated using the non-equilibrium Green's functions
\begin{equation}
%\tilde{G}(E)=\left((E+i\eta) I-H-\Sigma_{L}(E)-\Sigma_{R}(E)\right)^{-1},
\tilde{G}(E)=\left(EI - H-\Sigma_{L}(E)-\Sigma_{R}(E)\right)^{-1},
\label{eq:green}
\end{equation} 
where $\tilde{G}$ is Green's function, $E$ is the energy, $I$ is the identity matrix, and $\Sigma_{L(R)}$ is the self-energy associated with the semi-infinite left (right) lead. We obtain the self-energies in a self-consistent manner as
\begin{equation}
\Sigma_{L(R)}(E)=H_{1}^{\dagger}(EI-H_{0}-\Sigma_{L(R)}(E))^{-1}H_{1},
\end{equation}
with $H_{0}$ being the Hamiltonian of the unit cell in the lead and $H_{1}$  their coupling. With the self-energies at hand, the broadening function $\Gamma_{L(R)}$ is next calculated 
\begin{equation}
\Gamma_{L(R)}(E)=i[\Sigma_{L(R)}(E)-\Sigma_{L(R)}(E)^{\dagger}].
\end{equation}
Finally, the conductance $G(E)$ (in units of conductance quantum $G_{0}$) is achieved through the Landauer formula
\begin{equation}
G(E)=G_{0}T(E)=\dfrac{2e^{2}}{h}T(E),
\end{equation}
where the transmission coefficient $T(E)$ is given by 
\begin{equation}
T(E)=\textnormal{Tr}[\Gamma_{L}\tilde{G}\Gamma_{R}\tilde{G}^{\dagger}].
\end{equation}
All our calculations are performed with the \textsc{kwant} package \cite{KWANT}.

 \begin{figure}[t!]
  \centering
 \includegraphics[width=1\columnwidth]{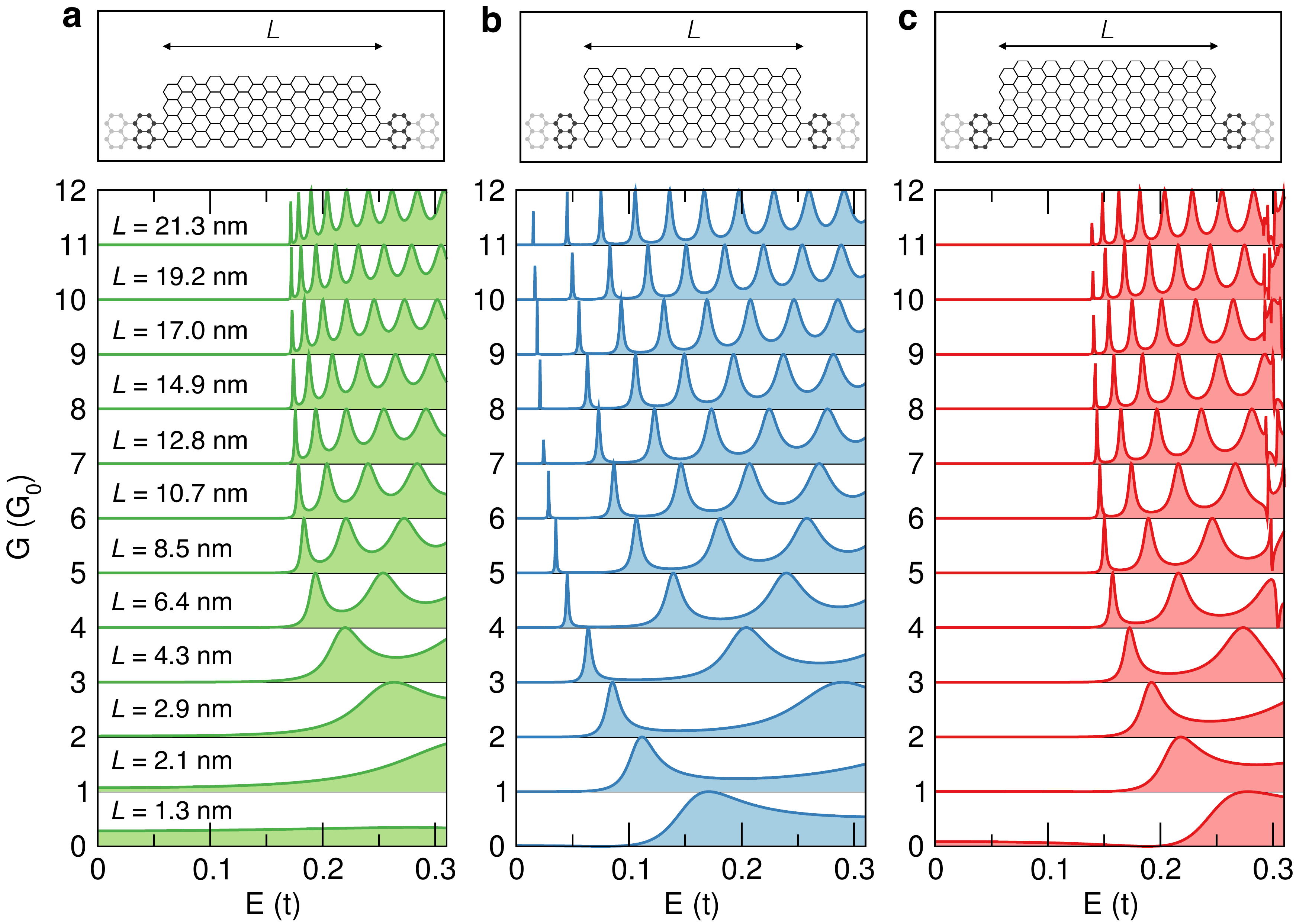}
  \caption{Atomic structure (upper panels) and conductance spectrum (lower panels) of (a) 5-10-5, (b) 5-11-5, and (c) 5-12-5-AGNR heterojunctions for increasing lengths of the quantum dot $L$, as indicated. Fermi level level is set to zero.  \label{Fig2}}
\end{figure}

 \begin{figure*}[t]
  \centering
 \includegraphics[width=2\columnwidth]{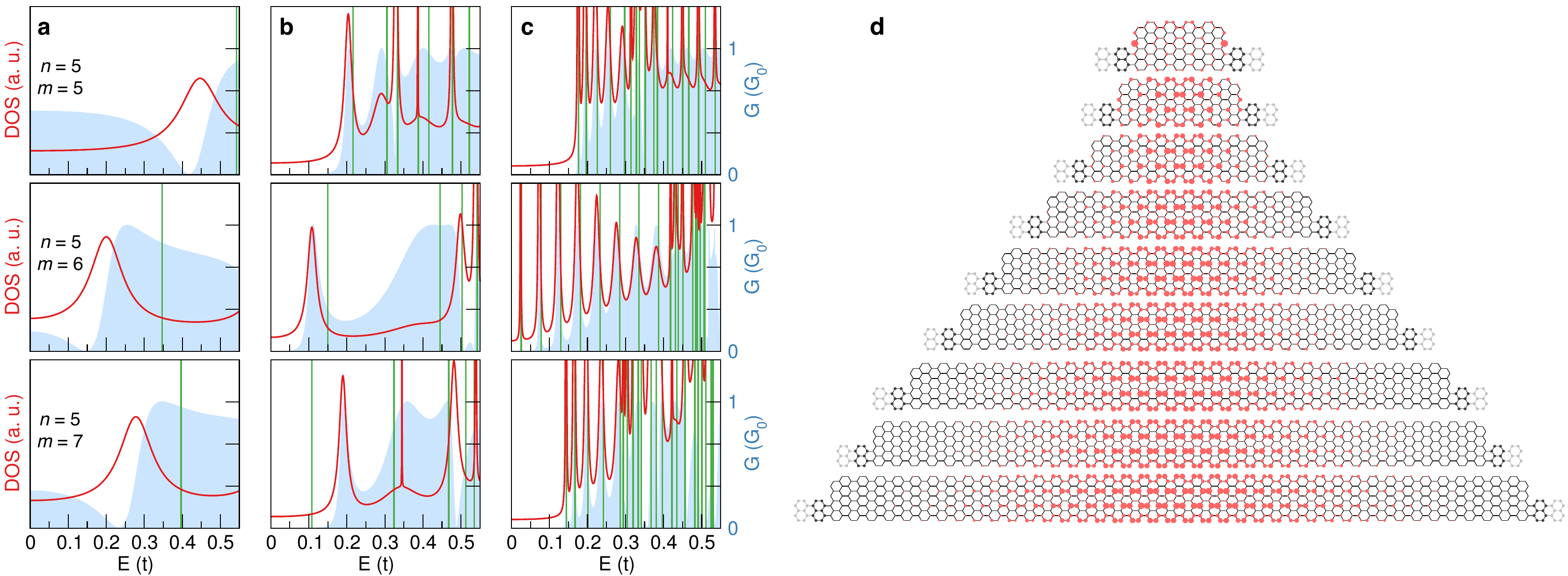}
  \caption{Selected representative density of states of the central channel (red lines), transmission spectrum (blue area), and energy levels of the isolated graphene quantum dot of width ($m + n$) (green lines) corresponding  to the (a) antiresonant ($L = 0.85$ nm), (b) intermediate ($L = 5.1, 3.0, 2.1$ nm for top, middle, and bottom panel, respectively), and (c) resonant tunnelling ($L = 12.8$ nm) electronic transport regimes discussed in the text for $n = 5$ and $m = 5$ (top panels), $m = 6$ (middle panels), and $m = 7$ (bottom panels). Fermi level is set to zero. (d) Evolution of the local density of states at the band edge with $L$ in the 5-10-5-AGNR heterojunction.   \label{Fig3} }
\end{figure*}

\medskip
\paragraph{Results and Discussion.} 
We consider the nanoarchitectures experimentally reported in \emph{e.g.}\ Ref.\ \cite{Wang2017, Jacobse2017}. As shown in Fig.\ \ref{Fig1}, our models are width-modulated AGNRs in which a channel containing a graphene quantum-dot of width $m$ and length $L$ is  contacted to a pair of equivalent semi-infinite leads of width $n$, hence giving rise to a $n$-$(n+m)$-$n$ in-plane heterojunction. As is customary, both $n$ and $m$ are quantified as the number of carbon atoms across the transport direction. We set $n = 5, 8, 11$ (that is, $n \in$ 3$p$ + 2) to ensure the metallic character of the leads at the adopted level of theory, whereas we extensively cover the width space of the quantum dot by choosing $5 \leq m \leq 13$. In analogy with experiments \cite{Wang2017}, we span a length interval of the quantum dot  $L = N \times a_0$ ranging from 0.9 to 21.3 nm, \emph{i.e.}\ corresponding to  $2 \leq N \leq 50$ unit cells of lattice constant $a_0 = 4.26$ {\AA}. The combination of the geometrical parameters $n$, $m$, and $L$ leads to over 1300 distinct atomic structures. For each of them, we have obtained both the conductance spectrum and the electronic density of states. These results are given in the Supplemental Material, which serves as an ``atlas" to understand the electronic properties and further guide the design of such width-modulated AGNRs \footnote{See Supplemental Material at [URL will be inserted by publisher] for additional results.}. Depending on the topological class to which each of the two constituent AGNRs belongs \cite{Cao2017}, some of the heterojunctions considered may feature topologically protected interface states. In Supplemental Fig.\ F1 \cite{Note1}, we compare the electronic properties of two structurally similar heterojunctions whose constituent pairs of AGNRs do (5-11-5-AGNR) or do not (5-9-5-AGNR) belong to distinct topological classes. Although the topologically protected state in the 5-11-5-AGNR heterojunction exhibits a strong localization at the interface between the channel and the lead, our results indicate that the transport gaps on which we focus in this work are not sensitive to it. In the following, we identify the general trends and uncover the physical effects that govern the electronic transport across these heterojunctions. Without loss of generality, we mainly focus on structures with $n=5$ and $m = 5, 6, 7$ and increasing $L$.

In Fig.\ \ref{Fig2}, we show the atomic structures of  5-10-5, 5-11-5, and 5-12-5-AGNR heterojunctions along with the evolution of their conductance spectra with the length of the  quantum dot. Due to the electron-hole symmetry encoded in the employed tight-binding Hamiltonian, only positive energies are given. Irrespectively of the width of the quantum dot, comparable changes are observed in the conductance spectra upon lengthening the channel region. Specifically, conductance peaks sharpen, become denser, and step in energy towards the Fermi level. From a qualitative point of view, this situation is analogous to that of a particle encountering a double (symmetric) rectangular potential barrier, when the separation between the barriers is widened.

On a more rigorous ground, in the vicinity of the band edge, we can identify a \emph{smooth} transition between two distinct transport regimes that emerge upon increasing the length of the graphene quantum dot. In Fig.\ \ref{Fig3}, we show the conductance spectrum and the density of states of the three selected heterojunctions mentioned above containing a short [Fig.\ \ref{Fig3}(a)], intermediate [Fig.\ \ref{Fig3}(b)], and long [Fig.\ \ref{Fig3}(c)] quantum dot. Also shown are the energy levels of the corresponding finite-size \emph{isolated} $(m+n)$-AGNR of length $L$. On the one hand, the introduction of a short (\emph{e.g.}\ approximately $L < 5$ nm for 5-10-5-AGNR) quantum dot preserves the metallic character, substantially diminishes the conductance at the band edge, and gives rise to asymmetric Fano lineshapes whose minimum is located in energy quite far from the level associated with the corresponding isolated dot \cite{Mendoza08, Deng14}. Such antiresonances closely resemble those previously observed in the case of graphene nanoribbons upon functionalization with a $p$-polyphenyl or polyacene functional groups to the edge \cite{Cernevics2020}. On the other hand, the introduction of a long (\emph{e.g.}\ approximately $L > 10$ nm for 5-10-5-AGNR) quantum dot turns the electronic transport into a resonant tunnelling regime, where a finite transport gap opens and conductance peaks discretize, assume a unitary value, and eventually match in energy the spectrum of the isolated  dot \cite{Xiong11}. These two transport regimes are bridged by an intermediate situation, where both resonant and antiresonant features coexist, though at different energy scales \cite{Xiong2011}. 

We suggest that the origin of the transition crossing the two transport regimes traces back to the different coupling strength between the states of the channel and those of the lead that occurs when moving from short to long quantum dots. This is further supported by the evolution of the local density of states (LDOS) at the band edge with the length of the quantum dot, as displayed in Fig.\ \ref{Fig3}(d). For short quantum dots, the LDOS localizes at the ends, thereby offering electronic states that are available to interact with those of the leads, hence promoting a strong channel-to-lead coupling.  Oppositely, long quantum dots host a LDOS that resides in the inner region solely, hence behaving rather independently from the contacts. In addition, we remark that, due to the different behavior in the antiresonant and resonant regimes, the difference in energy between conductance peaks and corresponding levels of the isolated quantum dot [$\delta$(E)] can be utilized as a suitable descriptor to monitor the transport crossover. This is indeed shown in the inset of Fig.\ \ref{Fig5}(f), where it can be noticed that the increase of $L$ rapidly decreases $\delta$(E), as we further detail below.

\begin{figure}[t]
  \centering
 \includegraphics[width=1\columnwidth]{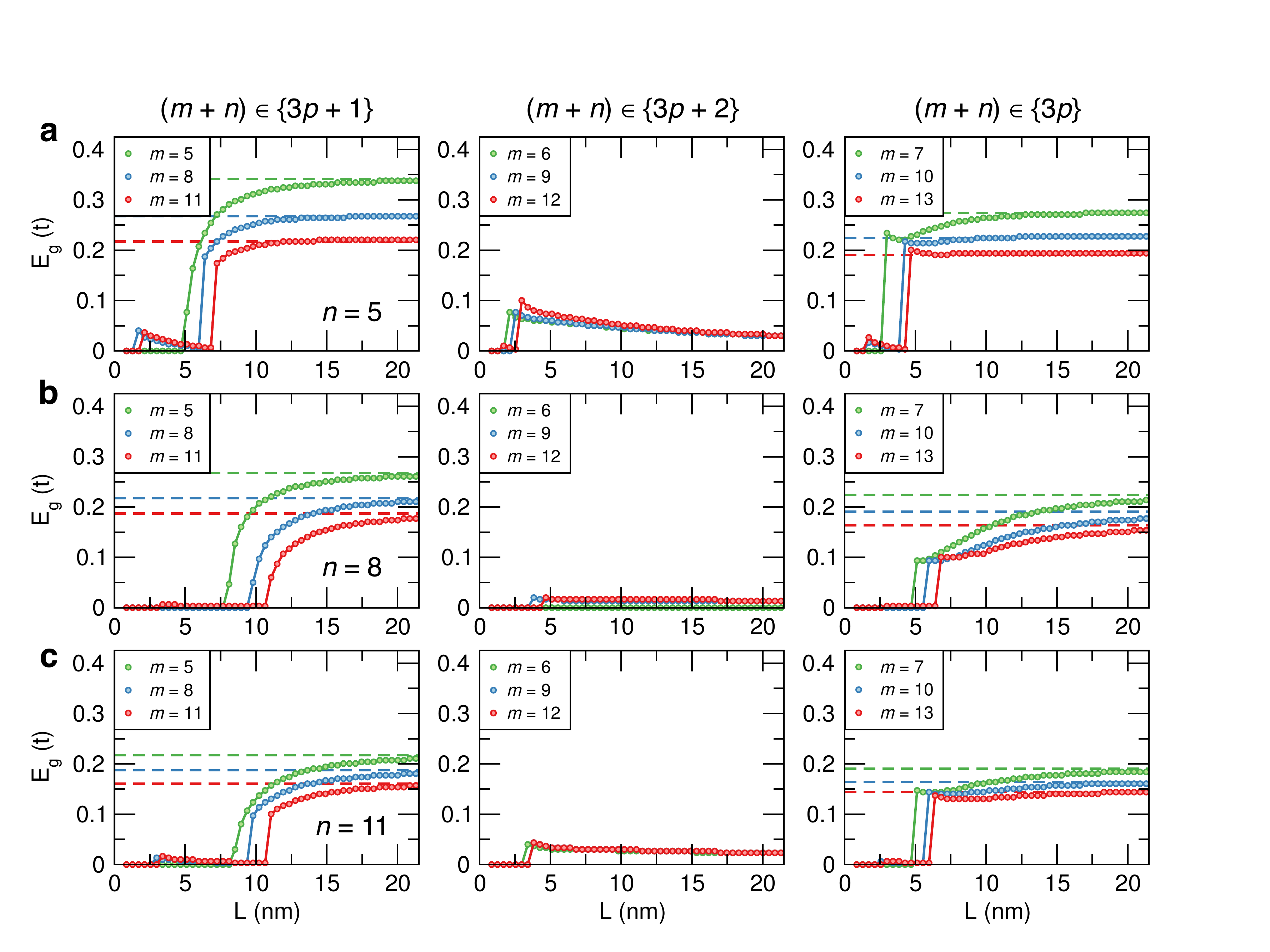}
  \caption{ Evolution of the transport gap $E\textsubscript{g}$ in $n$-$(m+n)$-$n$-AGNR heterojunctions as a function of the length of the quantum dot $L$ embedded in a lead of (a) $n = 5$, (b) $n = 8$, and (c) $n = 11$ and $5 \leq m \leq 13$. Dashed horizontal lines correspond to the band-gap of the corresponding periodic ($m+n$)-AGNR. \label{Fig4}}
  \end{figure}
  
  \begin{figure}[t]
  \centering
 \includegraphics[width=1\columnwidth]{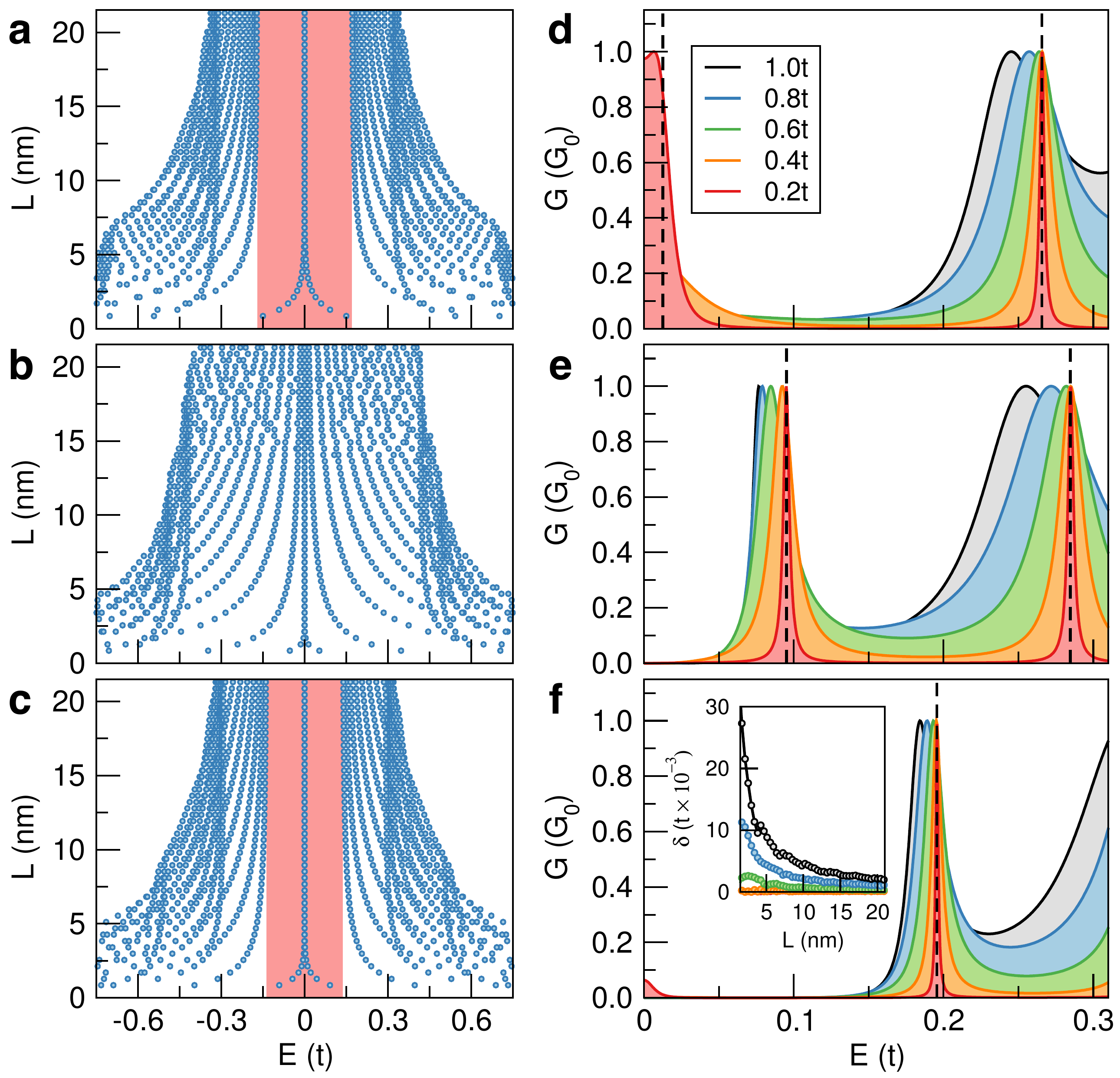}
  \caption{Energy spectra of finite-size (a) 10-, (b), 11-, and (c) 12-AGNR of a increasing length $L$. Red shaded area indicate the energy band-gap of the corresponding periodic AGNR. Evolution of the transmission spectrum with the hopping integral between the lead and the channel in (d) 5-10-5, (e) 5-11-5, and (f) 5-12-5-AGNR heterojunctions. The inset in panel (f) shows the energy shift between the energy levels of the isolated 12-AGNR and the resonant peaks in the transmission spectrum of the 5-12-7-AGNR junction as a function of $L$.
    \label{Fig5}}
\end{figure}

The weakening of the channel-to-lead coupling induced by the increase of $L$ and the accompanying transition between the two transport regimes discussed above have dramatic consequences on the transport gap. In Fig.\ \ref{Fig4}, we give such gap -- as extracted from the conductance spectra -- for all the heterojunctions considered in this work, featuring $n = 5$ [Fig.\ \ref{Fig4}(a)], $n = 8$ [Fig.\ \ref{Fig4}(b)], and $n = 11$ [Fig.\ \ref{Fig4}(c)]. Irrespective of the width ($m+n$) of the  channel, the introduction of short quantum dots and the ensuing antiresonant transport regime either retain the metallic character or open only a tiny transport gap. However, as the length of the quantum dot is increased and the resonant tunnelling regime is reached, a substantial transport gap opens and eventually converges to  that of the periodic $(m+n)$-AGNR. This implies that the  channel in the considered structures featuring either ($m+n$) $\in 3p$ or ($m+n$) $\in 3p + 1$ acquires semiconducting character \cite{Jacobse2017}, with a finite transport gap that can be engineered through the modulation of the quantum-dot width.  This finding indicates that these heterojunctions feature a semiconducting channel based on a graphene quantum-dot that is is seamlessly contacted to a pair metallic leads, thereby giving rise to complete all-carbon devices built into a single armchair graphene nanoribbon. Contrarily, the transport gap of heterojunctions featuring a quantum dot of witdth ($m+n$) $\in 3p+2$ vanishes with $L$.

The results reported in Fig.\ \ref{Fig4}, in which we show that the energy gap \emph{increases} with the length of the quantum dot, may appear counterintuitive on the basis of quantum-confinement effects solely, according to which one should expect the gap to \emph{decrease} with the length of the nanostructure \cite{Talirz2019}. In Fig.\ \ref{Fig5}(a-c), we present the energy spectra of finite-size AGNR of increasing length (notice that the zero-energy modes originate from the zigzag terminations and do not contribute to the electronic transport as they are selectively localized on one sublattice, hence effectively decoupled from the other lead). As expected, we observe that the energy gap decreases as the length of the AGNR increases, apparently at odds with the findings of Fig.\ \ref{Fig4}. However, we remark that another physical effect is operative and plays a central role in modulating the transport-gap width when the AGNR acts as a quantum dot. This corresponds to the coupling between the quantum dot (channel) and the semi-infinite AGNRs (lead), which in turn is very pronounced for small values of $L$, as we discussed above. In order to single out the impact of such coupling on the conductance spectrum, for the sake of illustration we gradually decrease the hopping integral $t$ in Eqn.\ (\ref{Ham}) between the atoms connecting the channel to the leads and obtain the resulting transmission spectrum \cite{Verg2018}, see Fig.\ \ref{Fig5}(a-c). Two distinct, though closely related, effects can be observed upon weakening the channel-to-lead coupling. Firstly, the conductance peaks step away from the Fermi level and approach the energy values of the isolated quantum dot AGNR, as shown in the inset of Fig.\ \ref{Fig5}(f). As the hopping integral decreases, this energy shift $\delta(E)$ vanishes, irrespective of the value of $L$ considered within the antiresonant transport regime as well.  Secondly, the broadening of the peaks in the conductance spectra is strongly reduced, such that the tails that were extending towards the Fermi level shrinks, with a metal-to-semiconductor transition takes place even for short $L$. Hence, the emergence of the two transport regimes (antiresonant \emph{vs.}\ resonant) and the consequent transport gaps (metallic  \emph{vs.}\ semiconducting) are found to be dominated by a subtle competition between the energy gap of the isolated quantum dot and the broadening of the conductance peaks that occurs upon contacting the channel to the leads.

\medskip
\paragraph{Summary and Conclusions.} We have investigated the electronic transport in width-modulated heterojunctions consisting of graphene quantum-dots embedded in metallic armchair nanoribbon leads, similar to those recently synthesized \emph{via} bottom-up approaches. We have considered over a thousand different configurations of varying geometries and determined their charge transport properties through a combination of atomistic tight-binding  and non-equilibrium Green's functions calculations. The emerging picture indicates that the conductance is found to be  dominated by the length scale of the quantum dot, which induces a smooth transition from a metallic antiresonant transport regime to a semiconducting resonant regime. Upon exceeding a critical length of the quantum dot -- the value of which is governed by the interplay between the intrinsic band-gap and the strength of the quantum dot-to-lead coupling -- a width-dependent transport gap opens, thereby giving rise to built-in one-dimensional metal-semiconductor-metal junctions.

To conclude, our work demonstrates that the experimentally realized functionalization of metallic AGNRs with graphene quantum-dots offer an effective route to \emph{directly} integrate a semiconducting channel into metallic electrical contacts while preserving the advantageous fine tunability of the AGNR band gap. Overall, our findings envisage graphene quantum-dot nanoarchitectures as self-contained electronic devices encoded in a sole graphene nanoribbon.

 \medskip
\paragraph{Acknowledgments.} The authors acknowledge the Swiss National Science Foundation (Grant No.\ 172543) and the NCCR MARVEL. M.P.\ is  financially supported by the Swiss National Science Foundation through the Early Postdoc.Mobility program (Grant No.\ P2ELP2-191706).  Calculations were performed at the Swiss National Supercomputing Center (CSCS) under the project s1008.

\end{document}